\documentclass[epj]{svjour}

\usepackage{graphicx}
\usepackage{amssymb}
\usepackage{color}

\newif\ifpreprint
\preprinttrue

\def\red{}
\begin{document}

\date{October 8, 2009}
\title{
\ifpreprint
\rightline{\normalsize UASLP--IF--09--001}
\vskip-0.3cm
\rightline{\normalsize FERMILAB--Pub--09--031--E}
\vspace{0.2cm}
\fi
Nuclear Dependence of Charm Production
}
\titlerunning{Nuclear Dependence of Charm Production}
\authorrunning{SELEX Collaboration, A.~Blanco-Covarrubias, et al.}


\author{
The SELEX Collaboration \\ 
A.~Blanco-Covarrubias\inst{11},
J.~Engelfried\inst{11}\mail{Jurgen Engelfried \email{jurgen@ifisica.uaslp.mx}},
U.~Akgun\inst{13},
G.~Alkhazov\inst{ 9},
J.~Amaro-Reyes\inst{11},
A.G.~Atamantchouk\inst{ 9}\fnmsep\thanks{deceased},
A.S.~Ayan\inst{13},
M.Y.~Balatz\inst{ 6}\fnmsep$^a$,
N.F.~Bondar\inst{ 9},
P.S.~Cooper\inst{ 4},
L.J.~Dauwe\inst{14}\fnmsep$^a$,
G.V.~Davidenko\inst{ 6},
U.~Dersch\inst{ 7}\fnmsep\thanks{Present address: Advanced Mask Technology Center, Dresden, Germany},
A.G.~Dolgolenko\inst{ 6},
G.B.~Dzyubenko\inst{ 6},
R.~Edelstein\inst{ 2},
L.~Emediato\inst{16},
A.M.F.~Endler\inst{ 3},
I.~Eschrich\inst{ 7}\fnmsep\thanks{Present address: University of California at Irvine, Irvine, CA 92697, USA},
C.O.~Escobar\inst{16}\fnmsep\thanks{Present address: Instituto de F\'{\i}sica da Universidade Estadual de Campinas, UNICAMP, SP, Brazil},
N.~Estrada\inst{11},
A.V.~Evdokimov\inst{ 6},
I.S.~Filimonov\inst{ 8}\fnmsep$^a$,
A.~Flores-Castillo\inst{11},
F.G.~Garcia\inst{16}\fnmsep\inst{ 4},
V.L.~Golovtsov\inst{ 9},
P.~Gouffon\inst{16},
E.~G\"ulmez\inst{ 1},
M.~Iori\inst{15},
S.Y.~Jun\inst{ 2},
M.~Kaya\inst{13}\fnmsep\thanks{Present address: Kafkas University, Kars, Turkey},
J.~Kilmer\inst{ 4},
V.T.~Kim\inst{ 9},
L.M.~Kochenda\inst{ 9},
I.~Konorov\inst{ 7}\fnmsep\thanks{Present address: Physik-Department, Technische Universit\"at M\"unchen, 85748 Garching, Germany},
A.P.~Kozhevnikov\inst{ 5},
A.G.~Krivshich\inst{ 9},
H.~Kr\"uger\inst{ 7}\fnmsep\thanks{Present address: The Boston Consulting Group, M\"unchen, Germany},
M.A.~Kubantsev\inst{ 6},
V.P.~Kubarovsky\inst{ 5},
A.I.~Kulyavtsev\inst{ 2}\fnmsep\inst{ 4},
N.P.~Kuropatkin\inst{ 9}\fnmsep\inst{ 4},
V.F.~Kurshetsov\inst{ 5},
A.~Kushnirenko\inst{ 2}\fnmsep\inst{ 5},
J.~Lach\inst{ 4},
L.G.~Landsberg\inst{ 5}\fnmsep$^a$,
I.~Larin\inst{ 6},
E.M.~Leikin\inst{ 8},
G.~L\'opez-Hinojosa\inst{11},
T.~Lungov\inst{16},
V.P.~Maleev\inst{ 9},
D.~Mao\inst{ 2}\fnmsep\thanks{Present address: Lucent Technologies, Naperville, IL},
P.~Mathew\inst{ 2}\fnmsep\thanks{Present address: Baxter Healthcare, Round Lake IL},
M.~Mattson\inst{ 2},
V.~Matveev\inst{ 6},
E.~McCliment\inst{13},
M.A.~Moinester\inst{10},
V.V.~Molchanov\inst{ 5},
A.~Morelos\inst{11},
A.V.~Nemitkin\inst{ 8},
P.V.~Neoustroev\inst{ 9},
C.~Newsom\inst{13},
A.P.~Nilov\inst{ 6}\fnmsep$^a$,
S.B.~Nurushev\inst{ 5},
A.~Ocherashvili\inst{10}\fnmsep\thanks{Present address: NRCN, 84190 Beer-Sheva, Israel},
Y.~Onel\inst{13},
S.~Ozkorucuklu\inst{13}\fnmsep\thanks{Present address: S\"uleyman Demirel Universitesi, Isparta, Turkey},
A.~Penzo\inst{17},
S.V.~Petrenko\inst{ 5},
M.~Procario\inst{ 2}\fnmsep\thanks{Present address: DOE, Germantown, MD},
V.A.~Prutskoi\inst{ 6},
B.V.~Razmyslovich\inst{ 9}\fnmsep\thanks{Present address: Solidum, Ottawa, Ontario, Canada},
V.I.~Rud\inst{ 8},
J.~Russ\inst{ 2},
J.L.~S\'anchez-L\'opez\inst{11},
J.~Simon\inst{ 7}\thanks{Present address: Siemens Healthcare, Erlangen, Germany},
A.I.~Sitnikov\inst{ 6},
V.J.~Smith\inst{12},
M.~Srivastava\inst{16},
V.~Steiner\inst{10},
V.~Stepanov\inst{ 9}\fnmsep$^m$,
L.~Stutte\inst{ 4},
M.~Svoiski\inst{ 9}\fnmsep$^m$,
N.K.~Terentyev\inst{ 9}\fnmsep\inst{ 2},
I.~Torres\inst{11}\fnmsep\thanks{Present address: Benemerita Universidad Aut\'onoma de Puebla (BUAP), Mexico},
L.N.~Uvarov\inst{ 9},
A.N.~Vasiliev\inst{ 5},
D.V.~Vavilov\inst{ 5},
E.~V\'azquez-J\'auregui\inst{11},
V.S.~Verebryusov\inst{ 6},
V.A.~Victorov\inst{ 5},
V.E.~Vishnyakov\inst{ 6},
A.A.~Vorobyov\inst{ 9},
K.~Vorwalter\inst{ 7}\fnmsep\thanks{Present address: Allianz Insurance Group IT, M\"unchen, Germany},
J.~You\inst{ 2}\fnmsep\inst{ 4},
R.~Zukanovich-Funchal\inst{16}  
}
\institute{
 Bogazici University, Bebek 80815 Istanbul, Turkey
 \and
 Carnegie-Mellon University, Pittsburgh, PA 15213, U.S.A.
 \and
 Centro Brasileiro de Pesquisas F\'{\i}sicas, Rio de Janeiro, Brazil
 \and
 Fermi National Accelerator Laboratory, Batavia, IL 60510, U.S.A.
 \and
 Institute for High Energy Physics, Protvino, Russia
 \and
 Institute of Theoretical and Experimental Physics, Moscow, Russia
 \and
 Max-Planck-Institut f\"ur Kernphysik, 69117 Heidelberg, Germany
 \and
 Moscow State University, Moscow, Russia
 \and
 Petersburg Nuclear Physics Institute, St.\ Petersburg, Russia
 \and
 Tel Aviv University, 69978 Ramat Aviv, Israel
 \and
 Universidad Aut\'onoma de San Luis Potos\'{\i}, San Luis Potos\'{\i}, Mexico
 \and
 University of Bristol, Bristol BS8~1TL, United Kingdom
 \and
 University of Iowa, Iowa City, IA 52242, U.S.A.
 \and
 University of Michigan-Flint, Flint, MI 48502, U.S.A.
 \and
 University of Rome ``La Sapienza'' and INFN, Rome, Italy
 \and
 University of S\~ao Paulo, S\~ao Paulo, Brazil
 \and
 University of Trieste and INFN, Trieste, Italy
}

\abstract{
Using data taken by SELEX during the 1996-1997
fixed target run at Fermilab, we study the production of charmed hadrons
on copper and carbon targets with $\Sigma^-$, $p$, $\pi^-$, and $\pi^+$
beams. Parametrizing the dependence of the inclusive 
production cross section on the atomic number $A$ as $A^\alpha$,
we determine $\alpha$
 for $D^+$, $D^0$, $D_s^+$, $D^+(2010)$,
$\Lambda_c^+$, and their respective anti-particles, as a function of their
transverse momentum $p_t$ and scaled longitudinal momentum $x_F$. 
Within our statistics there is no dependence of $\alpha$ on $x_F$
for any charm species for the interval $0.1 < x_F < 1.0$.  The average 
value of $\alpha$ for charm production by pion beams is 
$\alpha_{\rm meson}=0.850\pm 0.028$.
This is somewhat larger than the corresponding average 
$\alpha_{\rm baryon} = 0.755 \pm 0.016$ for charm production 
by baryon beams ($\Sigma^-$, $p$).
}

\maketitle

\section{Introduction}
The inclusive production of various outgoing hadrons from a specific incident
beam particle interacting in a nuclear target has been studied for many years.
The usual characterization of the process is that in a complex nucleus of
atomic number $A$, 
the single nucleon production cross section is increased by a 
factor $A^{\alpha}$.  
If the outgoing particle is absorbed with the same cross 
section as the incoming beam particle, then $\alpha \sim 2/3$.  Conversely,
production of heavy quarks by hard scattering with slow hadronization may lead
to minimal absorption and give $\alpha \sim 1$.

Experiments have used a range of techniques to determine $\alpha$ for a wide
selection of charm and strange hadrons.   Some experiments have studied
generic charm hadron production by detecting only muons from semi-leptonic
decays~\cite{Cobbaert:1986iw,Cobbaert:1988wz,Cobbaert:1988rf}. 
Others have measured $\alpha$ for hidden charm in $J/\psi$ 
production~\cite{Leitch:1999ea,Alessandro:2003pc,Abt:2008ya}.  
A third group have looked at specific charm or strange final
state hadrons~\cite{Heller:1977ku,Skubic:1978fi,Aleev:1986vm,Vecko:1988ia,Adamovich:1992fx,Alves:1992ux,Alves:1993jv,Leitch:1994vc,Adamovich:1996xf,Apanasevich:1997jd,Collaboration:2007zg}. 
Until this experiment, no single experiment covered the
whole family of charm hadrons over a wide range of kinematic variables $p_t$
and $x_F$, using $\pi^+$, $\pi^-$, $p$, and $\Sigma^-$ beams.  
For the first time
these new data allow one to look for systematic variations of $\alpha$ for
different kinematic regions, different groups of outgoing charm hadrons,
or different beam hadrons.  This broad data set is important for comparing to
the variety of theoretical models that have been proposed.

The production of strange particles with a proton beam 
shows~\cite{Heller:1977ku,Skubic:1978fi}
a strong dependence of $\alpha$ on both $x_F$ and $p_t$, with $\alpha<0.5$
in some bins of $(x_F,p_t)$.
Open charm
production was measured~\cite{Aleev:1986vm,Vecko:1988ia,Adamovich:1992fx,Alves:1992ux,Alves:1993jv,Leitch:1994vc,Adamovich:1996xf,Apanasevich:1997jd,Collaboration:2007zg}
for pion, proton, and neutron beams, in different ranges of $x_F$, 
for $D^\pm$, $D^0$, $D^*$, and $D_s$ mesons as well as the $\Lambda_c^+$,
but most experiments only had one beam particle and/or final state
hadron, or averaged over several mesons,
publishing just one value for $\alpha$.
In addition to the previously mentioned muon
measurements~\cite{Cobbaert:1986iw,Cobbaert:1988wz,Cobbaert:1988rf,Leitch:1999ea,Alessandro:2003pc,Abt:2008ya},
prompt neutrino production assumed to be from charm decays
is reported in~\cite{Duffy:1985hk}.
As summarized in~\cite{Vogt:2001nh,Lourenco:2006vw},
many measurements of
open charm production concentrate in the central region $x_F \sim 0$ and find
$\alpha \sim 1$ for the specific final states measured.  For charmonium
production or muon-triggered inclusive charm production at large $x_F$,
$\alpha$ decreases, approaching $2/3$.

Studying the basic production and suppression mechanisms in charm
hadroproduction is 
important for understanding non-perturbative aspects of heavy hadron 
production.  These results also impact other
fields like Heavy-Ion collisions (see a review~\cite{Frawley:2008kk})
and Cosmic Ray Physics, where air shower
Monte Carlo simulations have to take into account
the production of charm particles.

We present in this letter a new measurement for $\alpha$ in the range
of $0.1<x_F<1$, for 14~different open charm particles and decay modes,
produced by four different beam particles. 
In any single beam / charm-particle combination this 
experiment has statistics that are similar to or better than 
those from any other previous measurement.

\section{Experimental Apparatus}
The experimental setup of the SELEX experiment is described 
elsewhere~\cite{Russ:1998rr}.
We point out the
most important features  of the setup used in this analysis.  SELEX
is a 3-stage magnetic spectrometer, designed for high acceptance
forward ($x_F\gtrsim0.1$) interactions. $600\,\mbox{GeV}/c$ negative
($\simeq50\,\%$ $\Sigma^-$, $\simeq50\,\%$ $\pi^-$) 
and $540\,\mbox{GeV}/c$ positive beam particles
($\simeq92\,\%$ $p$, $\simeq8\,\%$ $\pi^+$),
interact in five target foils, described in Table~\ref{tab:targets}.
\begin{table*}
\sidecaption
\resizebox{0.7\hsize}{!}
{
\begin{tabular}{ r  c  c  c  l  c c }
\hline
\hline
Name & Material &  Thickness $L$ &  Position  &  \multicolumn{1}{c}{$A$} & 
Density $\rho$ &  $\lambda_{\rm int}$ \\ 
     &          &  [cm]  & [cm] &  &  [g/cm$^3$] & [$\%$]\\ 
\hline
\hline
S4 & Scintillator & 0.158 & \-7.27& -- & 1.03 & 0.20\\
6 & Copper & 0.159 & \-6.13 & 63.5 & 8.96 & 1.06 \\
7 & Copper & 0.119 & \-4.62 & 63.5 & 8.96 & 0.76 \\
8 &  Diamond & 0.220 & \-3.10 & 12 & 3.25 &  0.82 \\
9 &  Diamond & 0.220 & \-1.61 & 12 & 3.25 &  0.82 \\
10 & Diamond & 0.220 & \-0.11 & 12 & 3.25 &  0.82 \\
IC1 & Scintillator & 0.200 & 2.46 & -- & 1.03 & 0.25 \\
IC2 & Scintillator & 0.200 & 2.97 & -- & 1.03 & 0.25 \\
\hline
\hline
\end{tabular}
}
\caption
{Physical Properties of Materials in the Charm Production Targets
region. The layout is shown in Fig.~\ref{fig:targets}.}
\label{tab:targets}
\end{table*}
The physical properties of the target foils were measured before the
installation into the experimental setup, and the thicknesses and positions
were verified by measuring the positions of the primary vertices.
Every beam particle is individually tagged by a Transition Radiation Detector,
and the meson (baryon) contamination in the baryon (meson) beam is 
below $1\,\%$.

The spectrometer had silicon strip detectors to measure the incoming
beam and outgoing
tracks.
Momenta of particles deflected by the analyzing magnets were 
measured by a system of proportional wire chambers (PWCs), drift chambers and
silicon strip detectors.  Momentum resolution for a typical
$100\,\mbox{GeV}/c$ track
was $\sigma_p/p \approx 0.5\,\%$. 
Charged particle identification was performed with a Ring
Imaging Cherenkov detector (RICH)~\cite{Engelfried:1998tv}, which 
distinguished $K^{\pm}$ from $\pi^{\pm}$ up to $165\,\mbox{GeV/}c$.
The proton 
identification efficiency was $>95\,\%$ above proton threshold
($\approx 90\,\mbox{GeV}/c$). 
For pions reaching the RICH detector,
the total mis-identification probability
due to all sources of confusion was $<4\,\%$.

\begin{figure}[htb]
\includegraphics[width=\hsize]{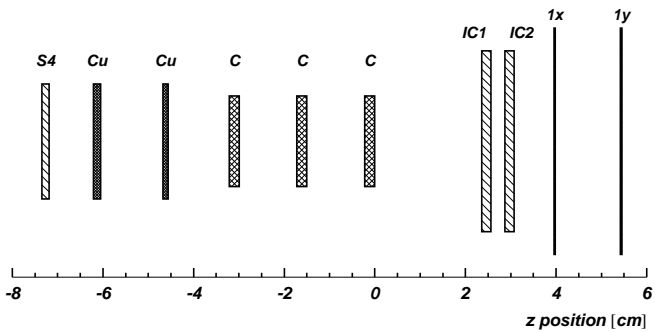}
\caption{Scale drawing of the charm production target region. In addition
to the five targets (2~Copper, 3~Diamond) we also indicate the location of
some of
the scintillators used in the trigger (S4, IC1, IC2) and the first two planes
(1x, 1y, from 20 in total) of the silicon strip detectors.
The physical properties
of the elements are shown in Table~\ref{tab:targets}.}
\label{fig:targets}
\end{figure}

Interactions in the five target foils were selected by a 
scintillator trigger.
The trigger for charm
required at least four charged tracks
downstream of the targets as indicated by an
interaction counter (IC1, IC2, see Fig.~\ref{fig:targets}), no signal
in a veto counter (S4) upstream of the targets,
 and at least 2 hits in a scintillator hodoscope after the
second analyzing magnet.  It accepted about 1/3 of all inelastic interactions.
Triggered events were further tested in an on--line computational filter 
based on downstream tracking and particle identification information.
The on--line filter selected events that had evidence of a secondary vertex
from tracks completely reconstructed using the forward PWC spectrometer and
the vertex silicon.  This filter reduced the 
data size by a factor of nearly~8 at a cost of about a factor of~2 in 
charm yield.
From a total of $15.2\cdot10^9$ interactions  
during the 1996--1997 fixed target run about $10^9$ events were
written to tape.

\section{Data Analysis}

To determine the charm production cross section dependence 
on the nuclear mass $A$, we have to determine the number of charm
particles produced in any single target, and take into account the
number of nuclei in the Carbon and Copper targets.
Parametrizing the cross section $\propto A^\alpha$, we obtain
\begin{equation}\label{eqn:first2}
\alpha= \frac{\ln \left( \frac {N_{\rm Cu}} {N_{\rm C}} 
                         \frac {\rho_{\rm C}}    {\rho_{\rm Cu}}
                         \frac {L_{\rm C}}       {L_{\rm Cu}}
                         \frac {A_{\rm Cu}}{A_{\rm C}}  \right)} 
             {\ln        \left( \frac {A_{\rm Cu}}      {A_{\rm C}}  \right)}
= 
{{\ln{{N_{\rm Cu}}\over{N_{\rm C}}}}\over{\ln{{A_{\rm Cu}}\over{A_{\rm C}}}}}
 +
\frac{\ln \left(
                         \frac {\rho_{\rm C}}    {\rho_{\rm Cu}}
                         \frac {L_{\rm C}}       {L_{\rm Cu}}
                         \frac {A_{\rm Cu}}{A_{\rm C}}  \right)} 
             {\ln        \frac {A_{\rm Cu}}      {A_{\rm C}} }
\end{equation}
with atomic masses $A_{\rm C}$, $A_{\rm Cu}$,
the thicknesses $L_{\rm C}$, $L_{\rm Cu}$, 
and densities $\rho_{\rm C}$, $\rho_{\rm Cu}$ 
as shown in Table~\ref{tab:targets}, and 
$N_{\rm C}$, $N_{\rm Cu}$ being the number of acceptance corrected
events observed in the different target materials.
This expression for $\alpha$ emphasizes the measurement issues.  The first
term depends on the corrected number of events from each target.  This means
that the acceptance and trigger efficiency for each different target have to
be understood as a function of the kinematic variables $x_F$ and $p_t$.  The
second term illustrates the requirement for precision in establishing the
parameters of each target used in the measurement.  Any uncertainties in
density or thickness translate directly into uncertainty or systematic
shift in $\alpha$.

In this analysis, we reconstructed completely
charm particles in specific decay modes.
For 
$D^0\to K^-\pi^+$, 
$D^0\to K^-\pi^+\pi^+\pi^-$, 
$D^+\to K^-\pi^+\pi^+$, 
$D_s^+\to K^-K^+\pi^+$, 
$\Lambda_c^+\to pK^-\pi^+$, and the corresponding charge-conjugated modes,
we used cuts similar to those in previous
publications~\cite{Kushnirenko:2000ed,Garcia:2001xj,Kaya:2003xd}.
Secondary vertex reconstruction was attempted when 
the $\chi^2$ per degree of freedom for the fit of the ensemble
of charged tracks to a single primary vertex exceeded~4. 
All combinations of tracks
were formed for secondary vertices ($\chi_{\rm sec}^2<5$)
and 
tested against a reconstruction table that specified 
selection criteria for each charm decay mode.
Secondary vertices which occurred inside the volume of a target
were rejected.
The resolution of the primary vertex position is on average better than 
$300\,\mu\mbox{m}$ (depending slightly on the target foil),
less than the thickness of the target foils and much less
than the spacing between foils. This permits an unambiguous assignment
of the interaction to a specific target foil.
Additional identification criteria for the different decay modes required
that 
proton and kaon candidate tracks were identified 
by the RICH detector to be at least as likely as a pion. Additionally,
in the case
of $D_s^\pm\to K^+K^-\pi^\pm$ for the kaon tracks the kaon hypothesis 
had to be more likely than the proton hypothesis.
If a pion candidate track reached the RICH detector, we applied
as a loose requirement that it had to have 
a likelihood of at least $10\,\%$;
if the track failed to reach the RICH, the candidate was called a pion.
The separation between the primary and secondary vertices 
had to be greater than eight times its uncertainty, and the 
uncertainty itself less than $0.17\,\mbox{cm}$;
the reconstructed charm momentum vector had to point back to the
primary vertex,
and two of the daughter tracks had to have a miss distance with
respect to the primary vertex of more than $\sqrt{6}$ times its uncertainty.
For $D^\star$ states decaying into $D^0\pi^\pm$,
we required a reconstructed
$D^0$ within $\pm36\,\mbox{MeV}/c^2$ ($\pm\,3$~times the resolution)
of the nominal mass, and an additional pion from the primary vertex.
The approximate total yields for the different modes and beam particles
are shown in Table~\ref{tab:yields}.
\begin{table*}
\sidecaption
\resizebox{0.7\hsize}{!}
{
\begin{tabular}{rlcccc}
\hline\hline
\multicolumn{2}{c}{~} & \multicolumn{4}{c}{Beam Particle} \\
\multicolumn{2}{c}{Decay Mode} & 
\multicolumn{1}{c}{$\Sigma^-$} & \multicolumn{1}{c}{$\pi^-$} &
\multicolumn{1}{c}{$p$} & \multicolumn{1}{c}{$\pi^+$} \\
\hline\hline
 1 & $D^0\to K^-\pi^+$ & 
  $1176 \pm 38$ & $411 \pm 22$ & $245 \pm 16$ & $29 \pm 7$ \\
 2 & $\overline{D^0}\to K^+\pi^-$ &
$1740 \pm 52$ & $452 \pm 23$ & $437 \pm 24$ & $39 \pm 7$ \\
 3 & $D^0\to K^-\pi^+\pi^+\pi^-$ &
$1282 \pm 50$ & $467 \pm 26$ & $252 \pm 18$ & $47 \pm 6$ \\
 4 & $\overline{D^0}\to K^+\pi^-\pi^+\pi^-$ &
$1650 \pm 60$ & $488 \pm 29$ & $331 \pm 26$ & $73 \pm 9$ \\
 5 & $D^+\to K^-\pi^+\pi^+$ &
$1352 \pm 46$ & $361 \pm 23$ & $248 \pm 20$ & $42 \pm 7$ \\
 6 & $D^-\to K^+\pi^-\pi^-$ &
$2024 \pm 58$ & $555 \pm 27$ & $338 \pm 22$ & $56 \pm 9$ \\
 7 & $D^{*+}\to D^0 (K^-\pi^+)\pi^+$ &
$165 \pm 13$ & $48 \pm 7$ & $33 \pm 7$ & \multicolumn{1}{c}{--} \\
 8 & $D^{*-}\to\overline{D^0}(K^+\pi^-)\pi^-$ &
$331 \pm 20$ & $70 \pm 8$ & $65 \pm 8$ & \multicolumn{1}{c}{--} \\
 9 & $D^{*+}\to D^0(K^-\pi^+\pi^+\pi^-)\pi^+$ &
$235 \pm 15$ & $61 \pm 9$ & $58 \pm 9$ & \multicolumn{1}{c}{--} \\
10 & $D^{*-}\to\overline{D^0}(K^+\pi^-\pi^+\pi^-)\pi^-$ &
$446 \pm 21$ & $116 \pm 11$ & $80 \pm 10$ & \multicolumn{1}{c}{--} \\
11 & $D^+_s\to K^-K^+\pi^+$ &
$118 \pm 17$ & $62 \pm 11$ & \multicolumn{1}{c}{--} & \multicolumn{1}{c}{--} \\
12 & $D^-_s\to K^+K^-\pi^-$ &
$379 \pm 26$ & $91 \pm 12$ & \multicolumn{1}{c}{--} & \multicolumn{1}{c}{--} \\
13 & $\Lambda^+_c\to pK^-\pi^+$ &
$1130 \pm 39$ & $172 \pm 15$ & $240 \pm 16$ &  \multicolumn{1}{c}{--}\\
14 & $\overline{\Lambda^-_c}\to\overline{p}K^+\pi^-$ &
$313 \pm 34$ & $95 \pm 13$ & $42 \pm 9$ & \multicolumn{1}{c}{--} \\
\hline\hline
\end{tabular}
}
\caption{Raw yields (before applying any acceptance corrections)
for the charm particles and modes, for the different beam
particles, used in this analysis.
These yields were obtained fitting a Gaussian and a 
polynomial representing the background to the invariant mass distributions.}
\label{tab:yields}
\end{table*}
The invariant mass distributions were divided 
into groups for the primary interaction happening in one of the
target foils,
and further in different $x_F$-bins, and
in some cases also in bins of $p_t^2$. 
We used the sideband-subtraction technique
        to remove the background from the mass distributions.  The resultant
736~different yields are the primary data for measuring $\alpha$.

\begin{table*}
\sidecaption
\resizebox{0.7\hsize}{!}
{
\begin{tabular}{cccccc}
 \hline
 \hline
Beam & Mode & $\alpha$ & $\alpha$ & $\alpha$ &$\alpha$ \\
& & $0.1<x_F<0.2$ & $0.2<x_F<0.4$ & $0.4<x_F<0.6$&$x_F>0.6$\\
 \hline
 \hline
$\Sigma^-$ &~1
 & $0.75\pm0.07$  & $0.72\pm0.07$  & $0.48\pm0.25$  & --  \\
$\Sigma^-$ &~2
 & $0.80\pm0.05$  & $0.70\pm0.06$  & $0.98\pm0.18$  & $0.71\pm1.54$  \\
$\Sigma^-$ &~3
 & $0.52\pm0.18$  & $0.66\pm0.09$  & $0.57\pm0.22$  & $0.67\pm0.68$  \\
$\Sigma^-$ &~4
 & $0.47\pm0.19$  & $0.67\pm0.09$  & $0.80\pm0.17$  & $1.23\pm0.92$  \\
$\Sigma^-$ &~5
 & $0.75\pm0.09$  & $0.68\pm0.07$  & $0.33\pm0.27$  & --  \\
$\Sigma^-$ &~6
 & $0.80\pm0.08$  & $0.79\pm0.06$  & $0.74\pm0.13$  & $0.84\pm0.59$  \\
$\Sigma^-$ &~7
 & $0.86\pm0.24$  & $0.89\pm0.15$  & $0.57\pm0.31$  & --  \\
$\Sigma^-$ &~8
 & $0.63\pm0.19$  & $0.73\pm0.11$  & $0.74\pm0.20$  & --  \\
$\Sigma^-$ &~9
 & $0.43\pm0.45$  & $0.41\pm0.17$  & $0.88\pm0.17$  & --  \\
$\Sigma^-$ & 10
 & $0.80\pm0.21$  & $0.80\pm0.10$  & $0.84\pm0.14$  & $0.47\pm0.50$  \\
$\Sigma^-$ & 11
 & $1.10\pm0.38$  & $1.07\pm0.19$  & --  & --  \\
$\Sigma^-$ & 12
 & $0.99\pm0.35$  & $0.79\pm0.12$  & $0.87\pm0.16$  & --  \\
$\Sigma^-$ & 13
 & $0.70\pm0.20$  & $0.95\pm0.08$  & $0.90\pm0.10$  & $0.83\pm0.17$  \\
$\Sigma^-$ & 14
 & $1.32\pm0.25$  & $0.74\pm0.24$  & --  & --  \\
 \hline
$\pi^-$ &~1
 & $0.86\pm0.14$  & $0.82\pm0.11$  & $0.25\pm0.24$  & --  \\
$\pi^-$ &~2
 & $0.89\pm0.12$  & $0.78\pm0.11$  & $0.96\pm0.17$  & --  \\
$\pi^-$ &~3
 & $0.85\pm0.32$  & $0.87\pm0.12$  & $0.82\pm0.17$  & $1.00\pm0.18$  \\
$\pi^-$ &~4
 & $1.04\pm0.22$  & $0.88\pm0.15$  & $0.59\pm0.19$  & $0.75\pm0.29$  \\
$\pi^-$ &~5
 & $0.37\pm0.34$  & $0.70\pm0.13$  & $0.79\pm0.17$  & $1.27\pm0.37$  \\
$\pi^-$ &~6
 & $0.76\pm0.18$  & $0.75\pm0.10$  & $0.89\pm0.16$  & $0.83\pm0.21$  \\
$\pi^-$ &~7
 & $1.50\pm0.56$  & $1.17\pm0.33$  & $0.68\pm0.36$  & --  \\
$\pi^-$ &~8
 & $0.34\pm1.06$  & $1.11\pm0.24$  & $0.77\pm0.32$  & --  \\
$\pi^-$ &~9
 & $1.54\pm0.60$  & $0.95\pm0.25$  & $0.29\pm0.46$  & $1.00\pm0.38$  \\
$\pi^-$ & 10
 & --  & $0.99\pm0.22$  & $0.70\pm0.25$  & $1.01\pm0.22$  \\
$\pi^-$ & 11
 & --  & $0.42\pm0.49$  & $0.83\pm0.58$  & --  \\
$\pi^-$ & 12
 & --  & $0.87\pm0.24$  & $0.68\pm0.44$  & $1.52\pm1.04$  \\
$\pi^-$ & 13
 & $1.42\pm0.54$  & $1.08\pm0.18$  & $0.84\pm0.27$  & $0.80\pm0.37$  \\
$\pi^-$ & 14
 & --  & --  & $0.95\pm0.37$  & --  \\
 \hline
$p$ &~1
 & $0.56\pm0.18$  & $0.65\pm0.14$  & $0.76\pm0.56$  & --  \\
$p$ &~2
 & $0.77\pm0.12$  & $0.67\pm0.12$  & $0.32\pm0.41$  & --  \\
$p$ &~3
 & $0.77\pm0.34$  & $0.53\pm0.20$  & $0.74\pm0.27$  & --  \\
$p$ &~4
 & $0.61\pm0.35$  & $0.45\pm0.20$  & $0.37\pm0.54$  & $0.93\pm2.32$  \\
$p$ &~5
 & $0.50\pm0.26$  & $0.71\pm0.14$  & --  & --  \\
$p$ &~6
 & $0.94\pm0.18$  & $0.80\pm0.12$  & $1.02\pm0.26$  & --  \\
$p$ &~7
 & --  & $0.61\pm0.35$  & $0.47\pm0.66$  & --  \\
$p$ &~8
 & --  & $1.03\pm0.19$  & $0.35\pm0.65$  & --  \\
$p$ &~9
 & --  & --  & $0.48\pm0.48$  & --  \\
$p$ & 10
 & --  & $0.73\pm0.23$  & $0.22\pm0.46$  & --  \\
$p$ & 13
 & $0.51\pm0.77$  & $0.44\pm0.23$  & $0.79\pm0.22$  & $1.03\pm0.30$  \\
$p$ & 14
 & --  & $1.08\pm0.80$  & $0.78\pm0.55$  & --  \\
 \hline
$\pi^+$ &~1
 & $0.37\pm0.65$  & $1.08\pm0.29$  & $0.23\pm0.65$  & --  \\
$\pi^+$ &~2
 & $0.68\pm0.34$  & $1.23\pm0.22$  & $0.38\pm0.68$  & --  \\
$\pi^+$ &~3
 & --  & $1.00\pm0.41$  & $0.91\pm0.35$  & $1.15\pm0.88$  \\
$\pi^+$ &~4
 & --  & $0.74\pm0.32$  & $1.57\pm0.51$  & --  \\
$\pi^+$ &~5
 & $1.11\pm0.67$  & $0.43\pm0.38$  & $0.58\pm0.80$  & --  \\
$\pi^+$ &~6
 & $0.59\pm0.76$  & $0.90\pm0.25$  & $1.28\pm0.33$  & $1.03\pm0.60$  \\
 \hline
 \end{tabular}

}
\caption{$\alpha$-values for the different
charm particles and decay modes, for different beam particles.
Only statistical uncertainties are shown. We do not determine $\alpha$ for
a specific mode and/or bin of $x_F$ if
the number of observed events (after sideband-subtraction) is below~1.}
\label{tab:allalphas}
\end{table*}

\begin{figure}[htb]
\includegraphics[width=\hsize]{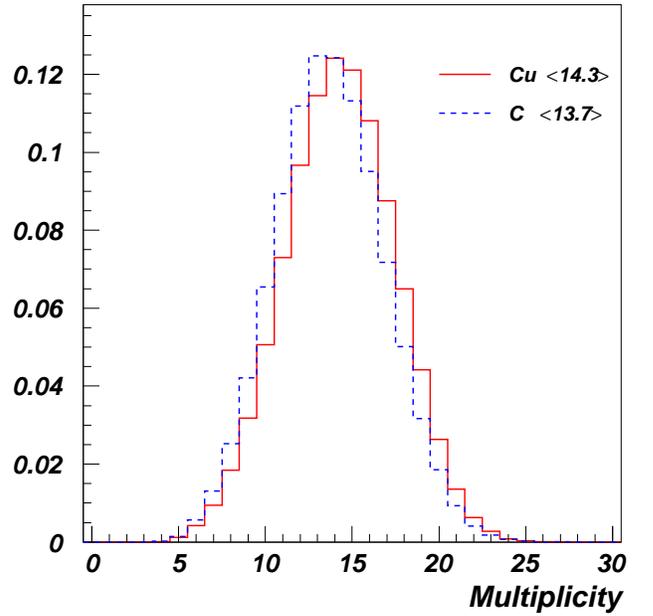}
\caption{{\red Multiplicity distributions
(number of tracks in the vertex region) for events with charm candidates
used in this analysis, for copper (full red) 
and carbon (dashed blue) targets. The corresponding averages are also shown.}}
\label{fig:multi}
\end{figure}

The total acceptance (geometrical acceptance and reconstruction efficiencies)
for the different decay modes of interest 
was estimated by embedding Monte Carlo charm decay
tracks into data events.  
Events were generated with 
in\-de\-pen\-dent\-ly-parametrized
transverse and longitudinal momentum distributions,
tuned to match the shapes of the data.
Detector hits, including resolution
and multiple Coulomb scattering smearing effects, produced
by these embedded tracks were folded into arrays of hits from real events.
The new ensemble of hits was passed through the SELEX off--line 
software.
We verified that the acceptance and reconstruction efficiency does not
depend on the multiplicity of the underlying event; additionally we found
that the multiplicity distributions are 
{\red nearly} independent of the target
material{\red, as shown in Fig.~\ref{fig:multi}}.
The acceptance is the ratio of the number of reconstructed events
over the number of embedded events in a particular mode for a specific target
foil 
and bin in $x_F$ and $p_t^2$.
As seen from equation~\ref{eqn:first2},
{\red most acceptance, trigger and filter effects cancel in this measurement,}
only differences in the acceptances between the target foils are important;
the largest effects depend on the lifetime of the different states.
For example, for $\Lambda_c^+$ decays in the $x_F$~interval $0.4$--$0.6$
the acceptance varies from
$21.4\,\%$ to $23.5\,\%$ between target~7 and target~10, while for
$D^+$ decays the variation is from $41.4\,\%$ to $31.8\,\%$.
We verified our acceptance corrections
by comparing the corrected event yields as functions of
$x_F$ and $p_t$ for our three identical diamond targets and found good
agreement within our statistics.
This study was performed for all the charm decay modes reported
here, as well as for the high statistics sample $\Lambda^0\to p\pi^-$.

For the determination of $\alpha$ we use the number of observed events,
corrected for acceptances, from the three diamond targets, but only
from the second copper target (target number~7). Some fraction of the
events with the interaction in target number~6 where vetoed in the
hardware trigger by the ``S4'' scintillation counter due to
back-splash from the interaction.  In the attempt to correct for this vetoing
we encountered systematic biases which would increase
the combined statistical and systematic uncertainties
more than
if we ignore completely the interactions from the first copper target foil. 
Nevertheless, the results using target 6 are consistent with those 
without it.

\section{Results}
For any single mode, for four different beam particles, we calculated 
$\alpha$ according to equation~\ref{eqn:first2}. The results are presented
in Table~\ref{tab:allalphas}.
{\red Since any single measurement has a large statistical uncertainty,
we average our results,
weighted by the individual uncertainties, 
for different groups of interest.}

\begin{figure*}
\hfill
\includegraphics[width=0.39\hsize]{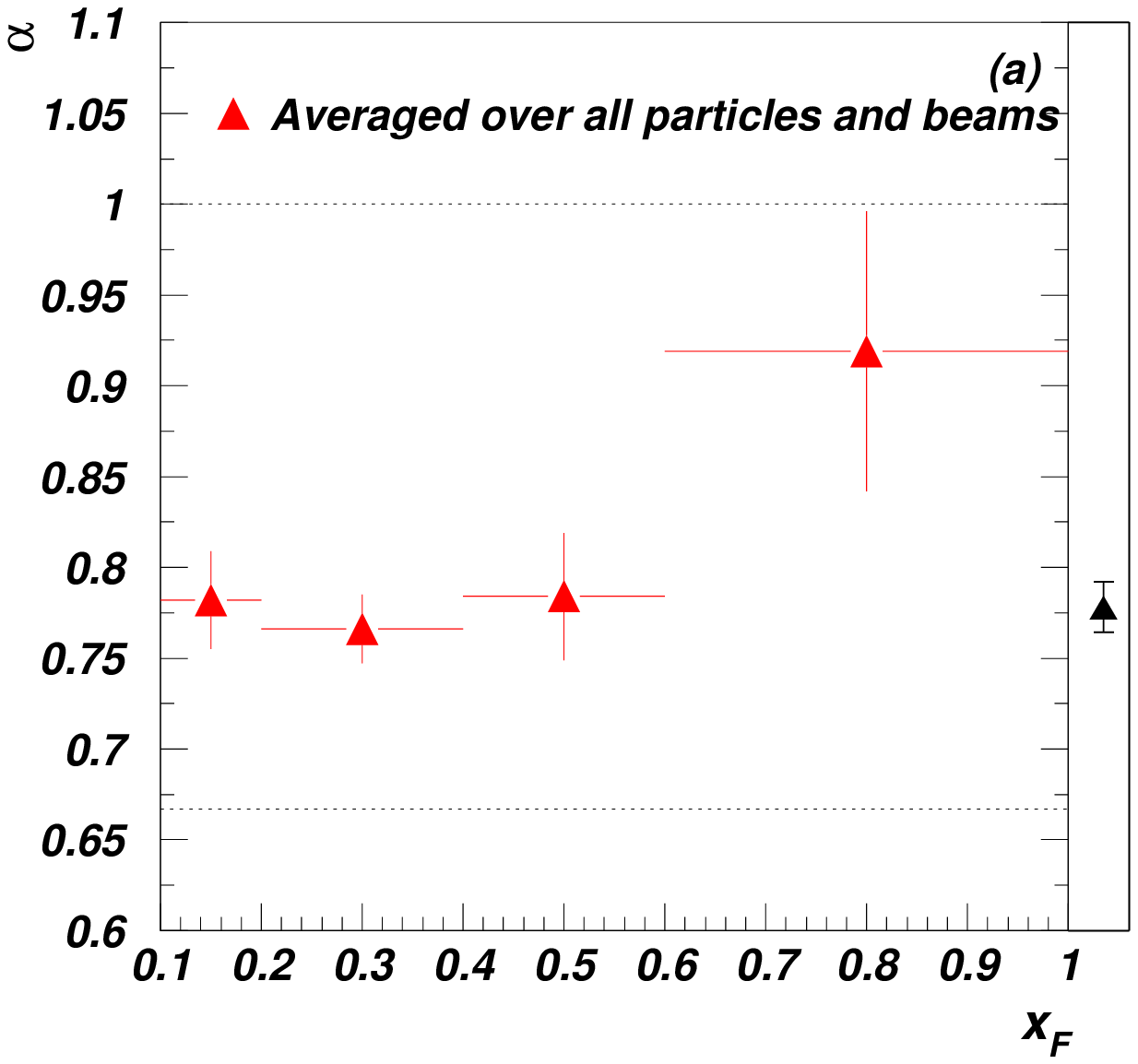}
\hfill
\includegraphics[width=0.39\hsize]{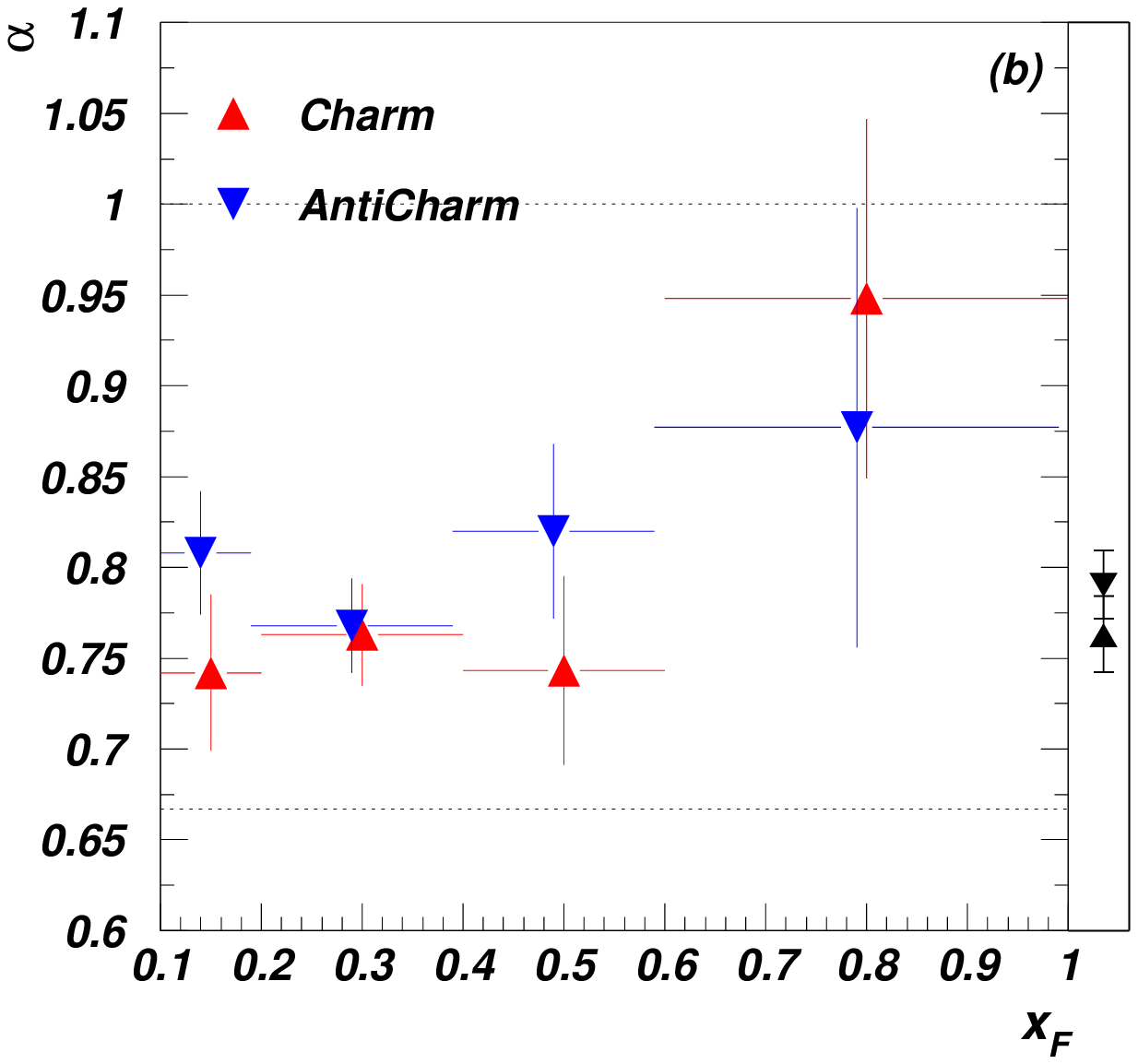}
\hfill{~}
\caption{Average $\alpha$ as function of $x_F$ 
for all observed final states (a) and 
for charm and anti-charm (b).
The data points are slightly offset to avoid overlapping of the error bars.
Reference $\alpha$ values of $2/3$ and $1$ are shown as dotted lines.
The points at $x_F>1$ show the
average assuming that $\alpha$ does not depend on $x_F$.
}
\label{fig:allcharm}
\end{figure*}

In Fig.~\ref{fig:allcharm}(a) we show $\alpha$ as a function of $x_F$ 
for all data, i.e.,\ averaged
over all charm and anti-charm modes and all beam particles.
In Fig.~\ref{fig:allcharm}(b) we 
separate the charm and anti-charm final states and show $\alpha$
averaged over 
all decay modes and beam particles for each type of charm quark.

\begin{figure*}
\includegraphics[width=0.33\hsize]{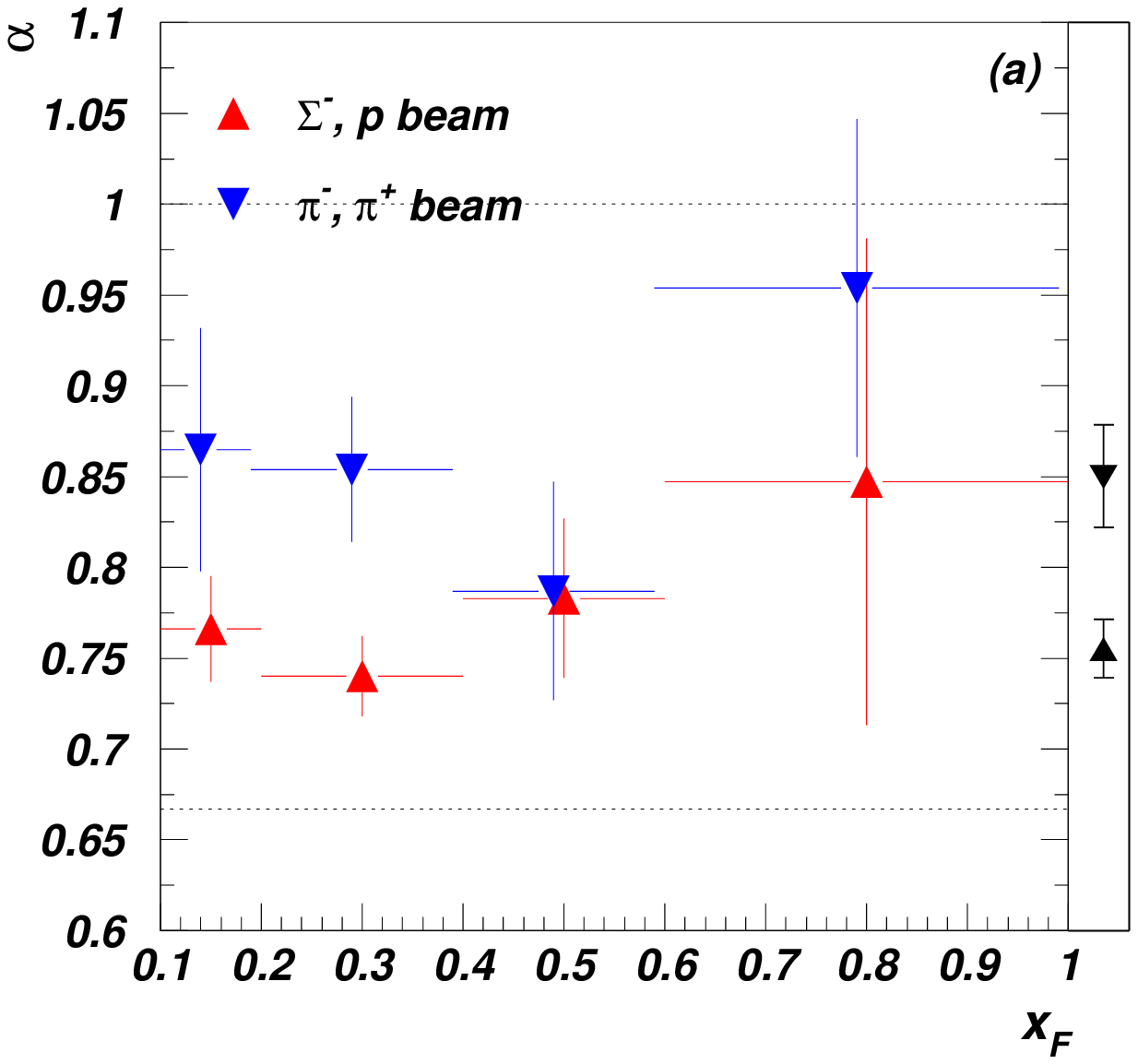}
\includegraphics[width=0.33\hsize]{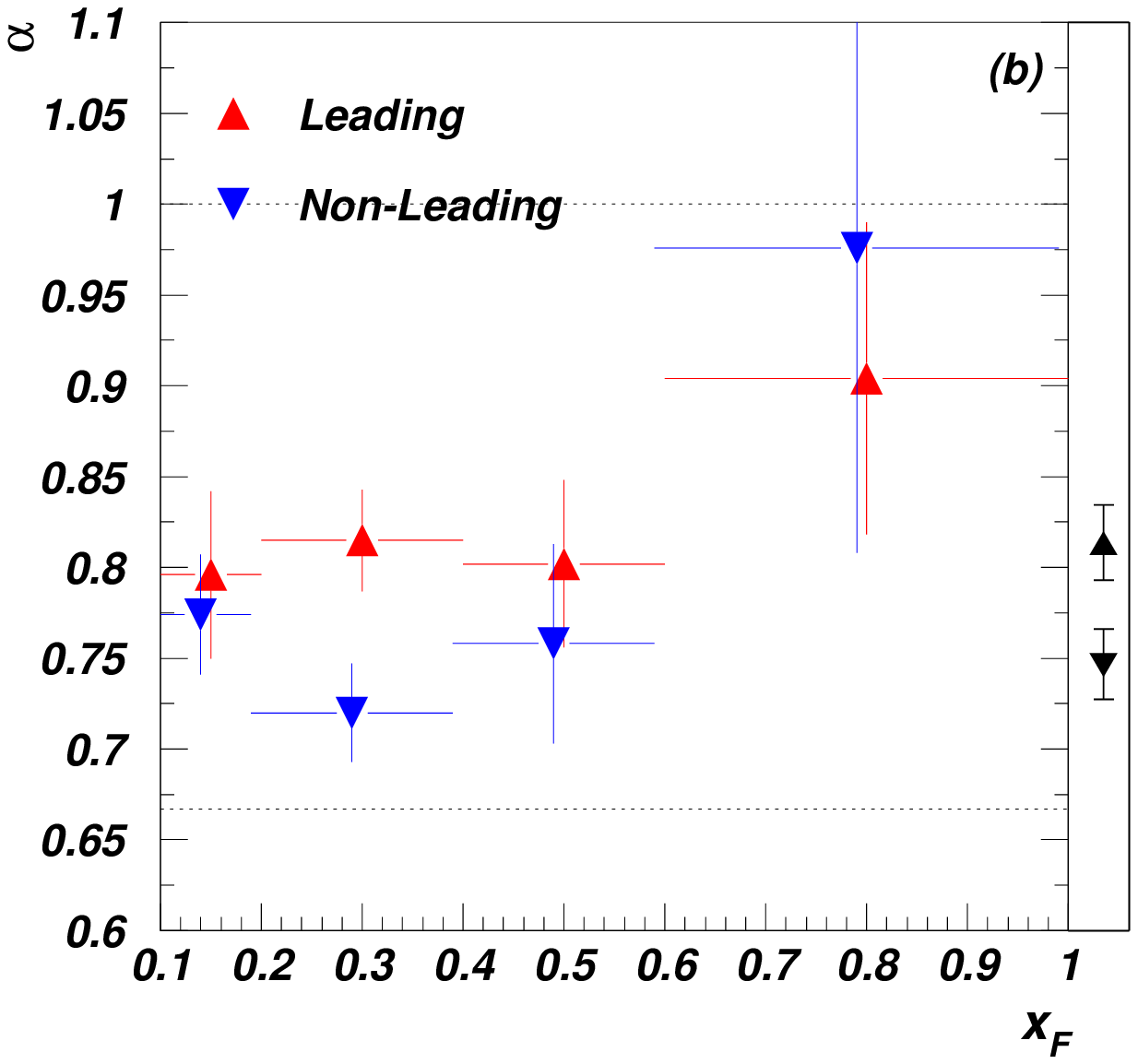}
\includegraphics[width=0.33\hsize]{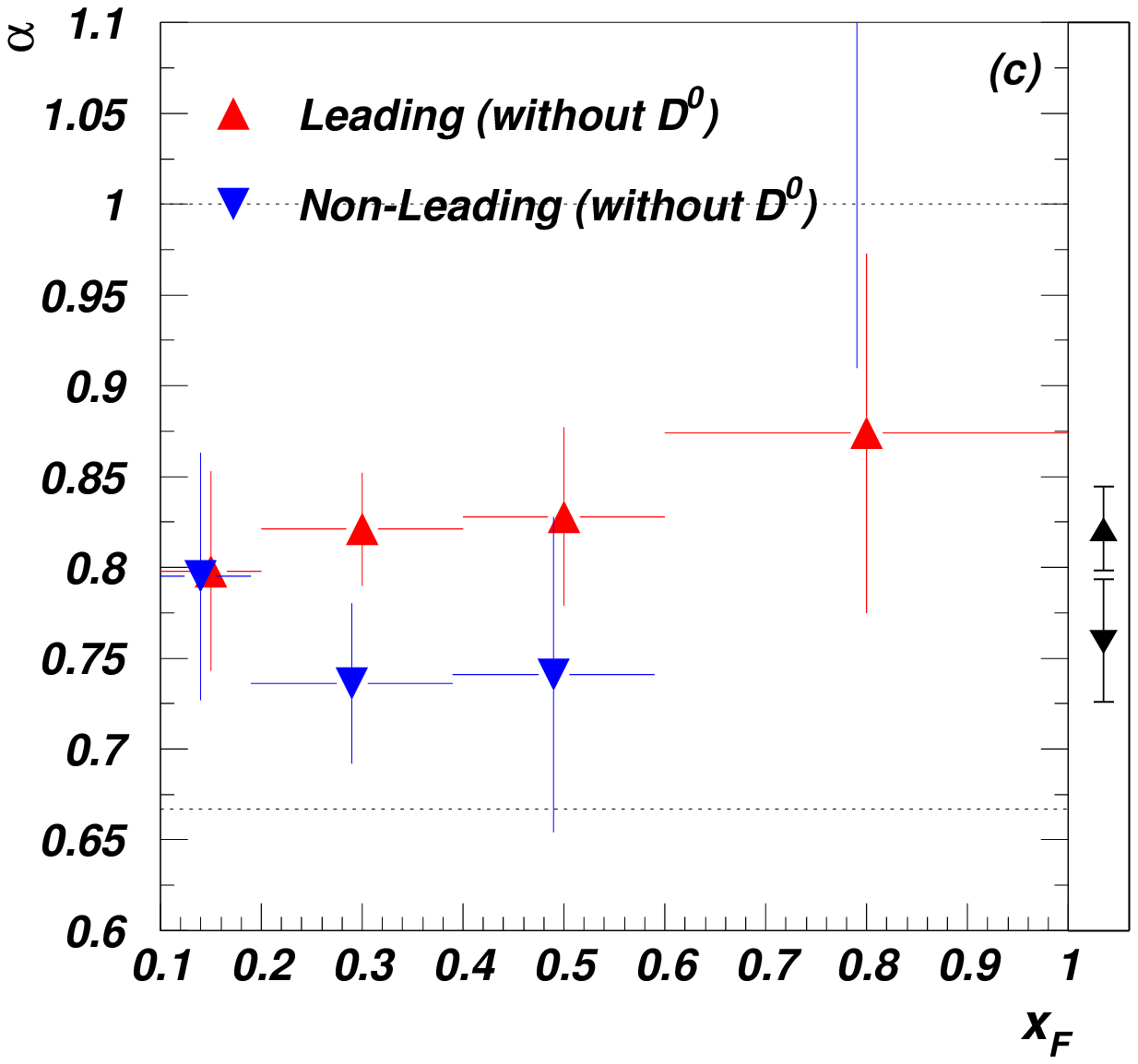}
\caption{Average $\alpha$ as a function of $x_F$ 
for production by baryon ($\Sigma^-$, $p$) 
and meson ($\pi^\pm$) beams (a) and
leading and non-leading (b{\red, c}) particles.
The data points are slightly offset to avoid overlapping of the error bars.
Reference $\alpha$ values of $2/3$ and $1$ are shown as dotted lines.
The points at $x_F>1$ show the
average assuming that $\alpha$ does not depend on $x_F$.
}
\label{fig:beamlead}
\end{figure*}

In Fig.~\ref{fig:beamlead}(a) we display the dependence of the 
average $\alpha$ on the beam
particle type: meson or baryon.  All charm and anti-charm decay modes are
averaged.  
In Fig.~\ref{fig:beamlead}(b{\red,c})
we separate the charm or anti-charm decays into 
leading and non-leading classes.  Recall that leading charm processes are
those in which the produced charm hadron carries at least one
valence quark of the beam
particle.  Non-leading charm processes have no valence quarks in common
between the beam and charm hadrons. 
{\red In Fig.~\ref{fig:beamlead}(c) we exclude the $D^0$ and 
$\overline{D^0}$ as these states could be produced preferably via excited 
states with different assignments to the leading/non-leading groups.}

\begin{figure*}
\hfill
\includegraphics[width=0.39\hsize]{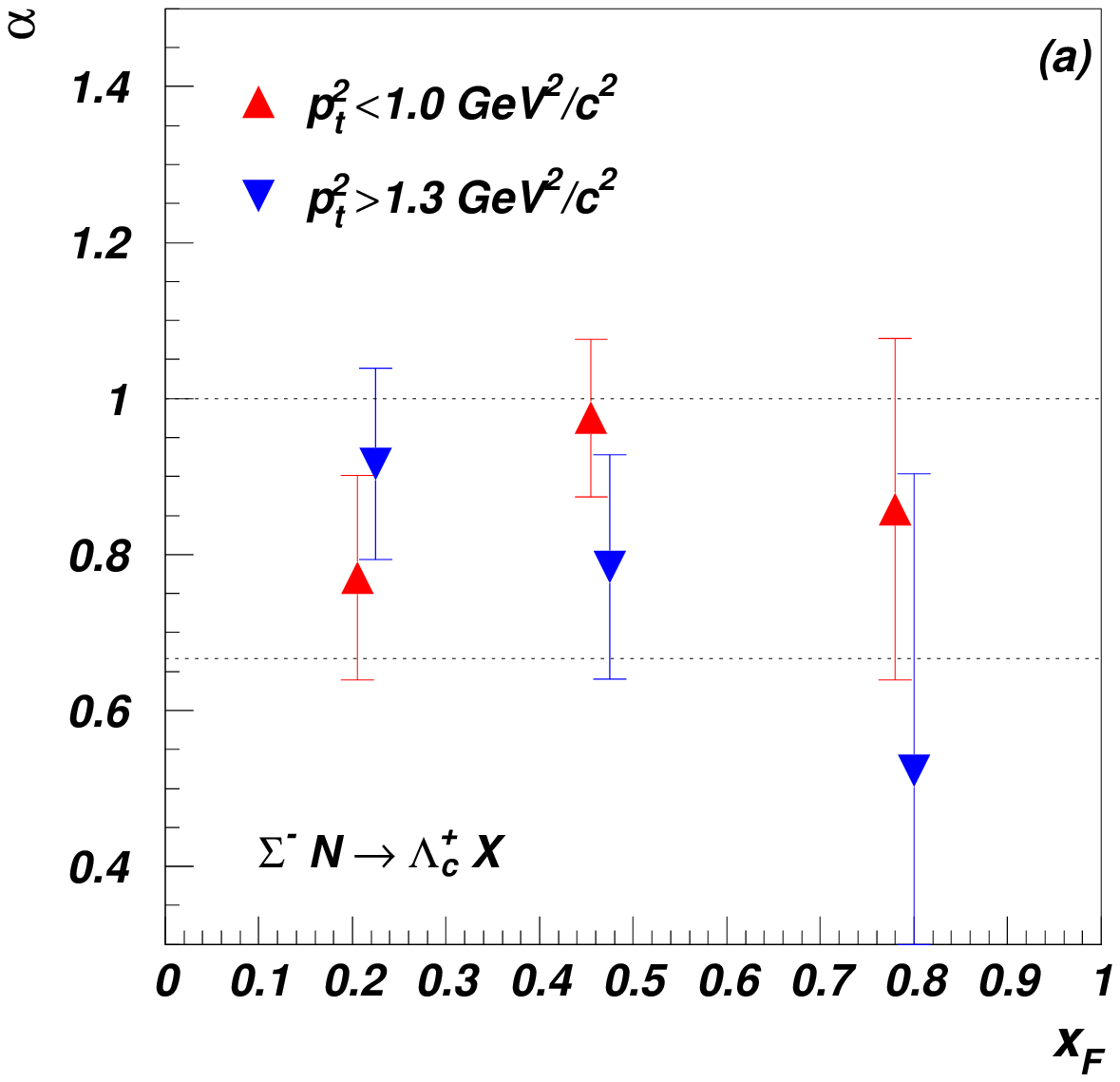}
\hfill
\includegraphics[width=0.39\hsize]{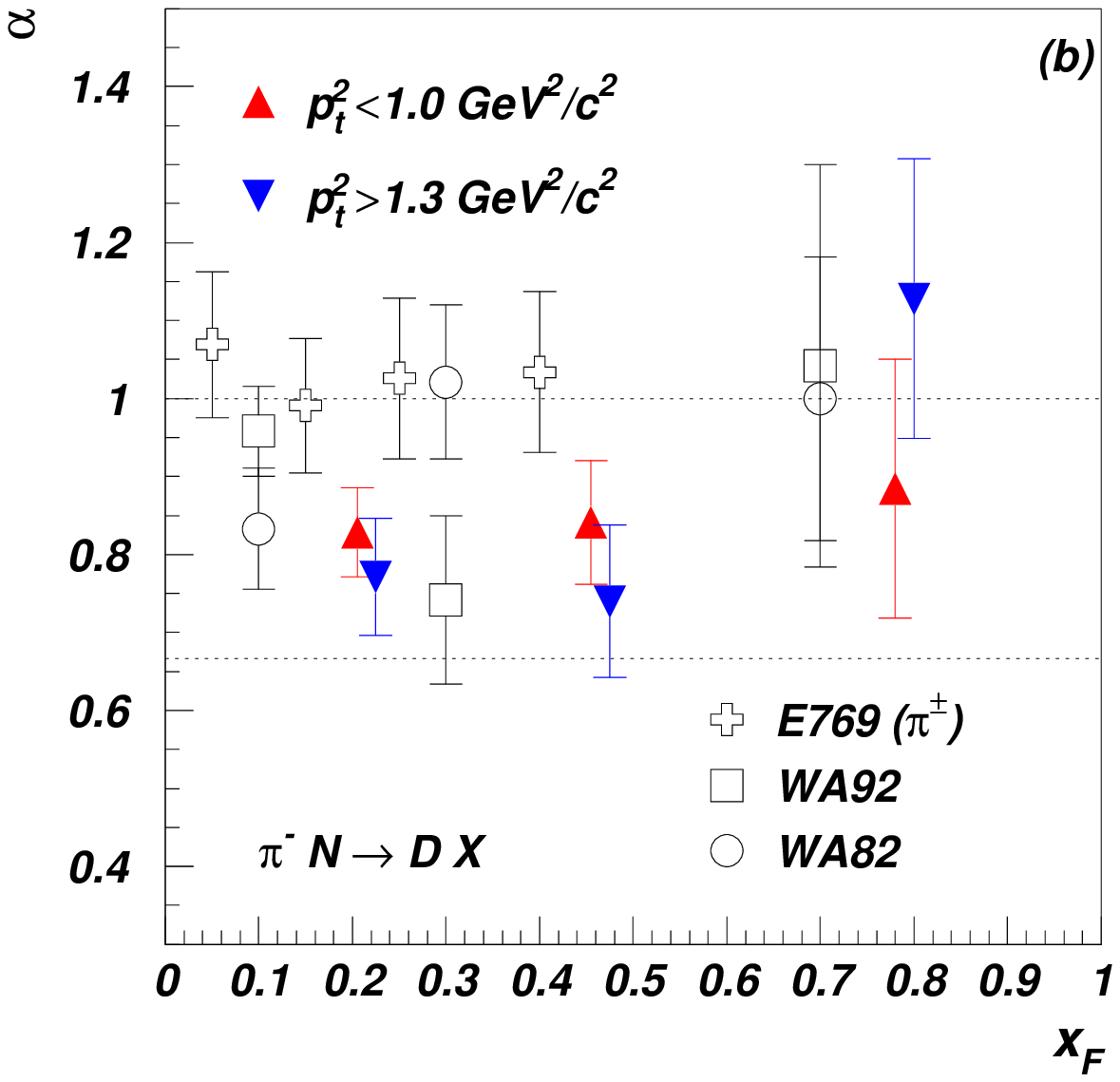}
\hfill{~}
\caption{$\alpha$ for the production of $\Lambda_c^+$ with a $\Sigma^-$ 
(a) and for $D^\pm$, $D^0$ mesons with $\pi^-$ (b) beam,
as function of $x_F$ for low ($p_t^2<1.0\,\mbox{GeV}^2/c^2$)
and high ($p_t^2>1.3\,\mbox{GeV}^2/c^2$)
transverse momentum.
Our data points are slightly offset to avoid overlapping of the error bars.
Reference $\alpha$ values of $2/3$ and $1$ are shown as dotted lines.
Also shown (open symbols) are
results~\cite{Adamovich:1992fx,Alves:1992ux,Adamovich:1996xf} from
other experiments, without separation in $p_t^2$.}
\label{fig:pt}
\end{figure*}

In Fig.~\ref{fig:pt}(a) we present the 
dependence of $\alpha$ on $x_F$ for low $p_t^2$ and high
$p_t^2$ events for $\Lambda_c^+$ production by
$\Sigma^-$ beam events, to look for
possible intrinsic charm effects~\cite{Brodsky:2006wb}.
Measurements of $\alpha$ are also 
available for $D$-meson production from the $\pi^-\,N$ 
experiments WA82~\cite{Adamovich:1992fx} {\red (Si, Cu, W targets)}, 
E769~\cite{Alves:1992ux} {\red (Be, Cu, Al, W)}, 
and 
WA92~\cite{Adamovich:1996xf} {\red (Cu, W)}. 
We compare our $D$-meson results from the $\pi^-$ data to
those from the other experiments in Fig.~\ref{fig:pt}(b).

As a systematic check, we
performed the identical analysis with $\Lambda^0$ and found good
agreement with previous measurements~\cite{Heller:1977ku,Adamovich:2003eu}
for both proton and $\Sigma^-$ beams.
Details will be 
presented in a forthcoming publication~\cite{alexlambda}.
We looked for variations of $\alpha$ with any of the
event selection cuts.  All changes were small compared with the statistical
uncertainty, indicating negligible systematic error from the cut selections.
We also studied binning effects and found only small shifts, compatible with
statistical uncertainties only.

\section{Discussion and Conclusions}

As seen from the figures, all the measured values are compatible with being
independent of $x_F$. Averaging over all our data, we obtain
$\alpha = 0.778\pm0.014$, which is incompatible with both usually suggested
values of $2/3$ and $1$. 
Additional averages for different groupings are show in
Table~\ref{tab:averages}.
\begin{table}
\caption{$\alpha$ values for various grouping of charm species and/or beams.}
\label{tab:averages}
\begin{tabular}{lc}
\hline\hline
\multicolumn{1}{c}{Average over $x_F$ and} & $\alpha$ \\ 
\hline\hline
all species and all beams & $0.778 \pm 0.014$ \\
all species for meson beams & $0.850\pm0.028$ \\
all species for baryon beams & $0.755\pm0.016$ \\
all beams for charm particles & $0.763\pm0.021$ \\
all beams for anti-charm particles & $0.791\pm0.019$ \\
all beams for leading particles & $0.814\pm0.021$ \\
{\red ~~~same as above, but excluding $D^0$'s} & {\red $0.821\pm0.023$} \\
all beams for non-leading particles & $ 0.747\pm0.019$\\
{\red ~~~same as above, but excluding $D^0$'s}& {\red $0.760\pm0.034$}\\
$\Lambda_c^+$ for $\Sigma^-$ beam, $p_t^2<1.0\,\mbox{GeV}^2/c^2$ & 
$0.894\pm0.075$\\
$\Lambda_c^+$ for $\Sigma^-$ beam, $p_t^2>1.3\,\mbox{GeV}^2/c^2$ & 
$0.841\pm0.091$\\
$D$ mesons for $\pi^-$ beam, $p_t^2<1.0\,\mbox{GeV}^2/c^2$ & 
$0.836\pm0.045$\\
$D$ mesons for $\pi^-$ beam, $p_t^2>1.3\,\mbox{GeV}^2/c^2$ & 
$0.796\pm0.057$\\
\hline\hline
\end{tabular}
\end{table}
Averaging separately over charm and anti-charm
final states the $\alpha$ values show no difference.
Separating into leading and non-leading production
(Fig.~\ref{fig:beamlead}(b)) we obtain
a $2.3\,\sigma$ difference{\red, which vanishes due to the 
increased uncertainties
when we exclude the $D^0$'s
in this comparison (Fig.~\ref{fig:beamlead}(c))}.
Separating the data into production
by meson beams and that by baryon beams (Fig.~\ref{fig:beamlead}(a)),
there is a
difference in the $\alpha$ value 
averaged over all charm and anti-charm modes,
corresponding to a $3\,\sigma$ effect.

For production of $\Lambda_c^+$ particles with a $\Sigma^-$ beam
the behavior shown in Fig.~\ref{fig:pt}(a) 
seems to suggest a decrease for high
$x_F$ and $p_t^2$, compared to $D$-mesons results shown in
Fig.~\ref{fig:pt}(b). 
We note that $D$-meson data has only a small contribution from
events with $p_t^2 > 1.3\,\mbox{GeV}^2/c^2$, 
so no firm conclusion can be drawn.  All
distributions are consistent with no dependence on $x_F$.

In summary, within our statistics there is no dependence of $\alpha$ on $x_F$
for any charm species for the interval $0.1 < x_F < 1.0$.  The average 
value of $\alpha$ for charm production by pion beams is 
$\alpha_{\rm meson} = 0.850\pm 0.028$.
This is somewhat larger than the corresponding average 
$\alpha_{\rm baryon} = 0.755\pm0.016$ for charm production 
by baryon beams ($\Sigma^-$, $p$).

\begin{acknowledgement}
We thank S.~Brodsky (SLAC) for useful discussions.
The authors are indebted to the staff of Fermi National Accelerator Laboratory
and for invaluable technical support from the staffs of collaborating
institutions.
This project was supported in part by Bundesministerium f\"ur Bildung, 
Wissenschaft, For\-schung und Technologie, Consejo Nacional de 
Ciencia y Tecnolog\'{\i}a {\nobreak (CONACyT)},
Conselho Nacional de Desenvolvimento Cient\'{\i}fico e Tecnol\'ogico,
Fondo de Apoyo a la Investigaci\'on (UASLP),
Funda\c{c}\~ao de Amparo \`a Pesquisa do Estado de S\~ao Paulo (FAP\-ESP),
the Israel Science Foundation founded by the Israel Academy of Sciences and 
Humanities, Istituto Nazionale di Fisica Nucleare (INFN),
the International Science Foundation (ISF),
the National Science Foundation (Phy \#9602178),
NATO (grant CR6.941058-1360/94),
the Russian Academy of Science,
the Russian Ministry of Science and Technology,
the Russian Foundation for Basic Research (grant 08-02-00657),  
the Secretar\'{\i}a de Educaci\'on P\'ublica (Mexico)
(grant number 2003-24-001-026),
the Turkish Scientific and Technological Research Board (T\"{U}B\.ITAK),
and the U.S.\ Department of Energy (DOE grant DE-FG02-91ER40664 and
DOE contract number DE-AC02-76CHO3000).
\end{acknowledgement}

\end{document}